# Robust topological BIC nanocavities for upconversion directional emission

Yongqi Chen[1], Ming Zhu[2], Qingfeng Bian[1], Xiumei Yin[3], Wenxin Wang[2, *], Bin Dong[3,*], and Yurui Fang[1, *]

[1.] *School of Physics, Dalian University of Technology, Dalian 116024, China.*
[2.] *Qingdao lnnovation and Development Center, Harbin Engineering University, Qingdao, 266500, China*
[3.] *School of Physics and Materials Engineering, Dalian Nationalities University, Dalian 116600, China*

*Corresponding authors: wenxin.wang@hrbeu.edu.cn (W.W), dong@dlnu.edu.cn (B.D.), yrfang@dlut.edu.cn (Y.F.)

**Abstract:**

Photonic bound states in the continuum (BICs) provide a revolutionary paradigm for boosting light-matter interactions in integrated nanocavity systems. Nevertheless, precise manipulation of open cavity-emitter architectures still faces critical challenges, especially in realizing deterministic directional radiation and suppressing the perturbation of intrinsic cavity modes induced by emitters as local impurities. Conventional investigations on cavity-emitter coupling are predominantly based on ensemble measurements, which inevitably mask the intrinsic physics underlying individual light-matter interactions. Here, we propose a robust strategy to control the upconversion and emission of a single-particle emitter using a topological plasmonic cavity with broken $\sigma_h$ mirror symmetry. This structured design enables the transition from symmetry-protected BICs to a multi-BIC regime with finite but ultrahigh confinement, where nontrivial phase evolution and hybridization of transverse electric and magnetic modes open a well-defined far-field radiation channel for directional emission. Leveraging this scheme, we experimentally demonstrate dramatically enhanced radiation intensity from a single point-like emitter, together with uniform and deterministic directional emission, while achieving excellent structural robustness against local perturbations. This work establishes a general framework for engineering coherent directional light emission at the nanoscale, which lays a solid foundation for high-performance chip-scale integrated nanophotonic applications.

**Introduction**

Spatial manipulation and directional control of nanoscale light emission lie at the core of next-generation nanophotonic technologies, ranging from nonlinear bioimaging[1–3], and single-photon sources to nanolasers[4–6], dynamic metasurface modulators and on-chip photonic routing [7–11]. Over the past decades, extensive efforts have been devoted to enhancing spontaneous emission and radiation efficiency through plasmonic nanocavities[12–16], whispering-gallery resonators[17–19] and surface lattice resonances[20–23]. While these platforms enable considerable light-matter interaction enhancement, they suffer from inherent limitations including fixed emission directionality governed by rigid mode dispersion, low quality factors, and severe mode distortion induced by local emitters acting as structural impurities[24–27]. These bottlenecks fundamentally restrict the realization of deterministic, robust, and high-efficiency directional emission required for high-precision imaging, holography, and multiplexed optical communication.

Topological photonics has emerged as a powerful framework to engineer robust light confinement and wave manipulation via nontrivial band topology[28–33], offering unprecedented immunity to structural defects and local perturbations. Among diverse topological photonic states, bound states in the continuum (BICs) stand out as ideal candidates for extreme light localization, featuring theoretically infinite quality factors and deeply subwavelength field confinement. By breaking specific spatial symmetries, symmetry-protected BICs can be converted into quasi-BICs (q-BICs) with finite yet ultrahigh quality factors, enabling strong radiation enhancement and controllable far-field output[34–39]. Despite rapid progress in BIC-enabled light manipulation across various integrated platforms [32,40–44], most existing studies focus on the collective response of emitter ensembles[16,25,45], which inevitably blurs the intrinsic photophysical mechanisms of single-emitter radiation and directionality at the individual particle level[46]. Although topological cavity-emitter systems have recently emerged as a promising route to address this challenge[47], to date, a universal, scalable, and low-cost platform that simultaneously achieves ultrahigh confinement, deterministic directional emission, and single-emitter compatibility in a robust topological cavity system remains largely unexplored.

Here, we theoretically design and experimentally realize a versatile platform for directional upconversion emission from a single nanocrystal (NC) by explicitly breaking the $\sigma_h$ horizontal mirror symmetry in a topological plasmonic cavity (TPC, Fig. 1a). Leveraging mask-free guided anodization and wet chemical etching of long-range ordered porous alumina templates, we fabricate large-area plasmonic nanoantenna arrays that drive the transition from symmetry-protected BICs to high-Q multi-BICs, opening well-defined far-field radiation channels (Fig. 1b). The engineered TPC strongly confines optical fields at the antenna tips, yielding a deeply subwavelength mode volume ($V_{\text{eff}}$). At the BIC resonance, we achieve an upconversion luminescence enhancement exceeding two orders of magnitude (147-fold), originating from the synergistic combination of ultrahigh Q-factor and accelerated radiative decay rates. The significantly boosted local density of optical states (LDOS), which spectrally overlaps with the 550 nm upconversion emission of single nanocrystals, leading to pronounced radiative enhancement. Through precise $\sigma_h$ symmetry engineering (Fig. 1c), we achieve deterministic control over control over the emission direction of single NCs, efficiently funneling upconversion photons into well-defined directional radiation cones (Fig. 1d) with exceptional structural robustness against local perturbations. By directly integrating individual nanocrystals into the topological plasmonic lattice, our platform overcomes the limitations of conventional ensemble-based systems and establishes a general route toward nanoscale coherent directional nonlinear light sources. This work opens new opportunities for high-performance chip-scale nanophotonic applications including 3D projection imaging, directional optical antennas, and augmented reality photonic waveguides.

## Results

### TPC symmetry breaking band engineering

The boosting directional system schematic in Fig. 1a illustrates a cavity-emitter composed of our TPC and a single-NC. The periodic TPC is fabricated from an aluminum film pre-patterned with a high-density square lattice of nanopores (Fig. 1b, left). A two-step anodization process[48] followed by wet chemical etching is employed to convert this nanopore array into well-defined topological plasmonic cavity (Fig. 1b, right and Supplementary Materials Fig. S2). This lithography-free method enables large-area fabrication of ordered nanostructures and is compatible with flexible electronics and nanophotonic applications. The resulting array, with a lattice constant of $a = 400$ nm, is shown in the SEM image in Fig. 1b. At the intersection points of the grid, aluminum nanocones are located, while intercell ridges are shown along the grid lines. In large-area TPC, monodisperse upconversion NCs (Fig. 1a and Supplementary Materials Fig. S4) were deposited onto the TPC, with single-NC lying horizontally filling the inter-cellular gaps, as shown in the inset of Fig. 1a.

To illustrate the above picture, we start from a standard photonic crystal template. When the incident light is reflected by the traditional flat periodic structure, its wavefront is partially shaped by the dispersion relation of the optical resonance. Specifically, the interaction between the incident light and the periodic structure can be decomposed into two processes: part of the energy is directly reflected by the TPC without coupling to the resonant mode, while the other part excites the resonance and radiates[39]. At resonance, the enhancement factor of the single resonance intensity in the open cavity can be described by the temporal coupled-mode[49]:

$$G = \frac{|E_{loc}|^2}{|E_i|^2} \approx \kappa_i^2 \frac{Q_{tot}^2}{V_{eff}} = \frac{Q_{tot}^2}{Q_r V_{eff}} = \frac{Q_r^2 Q_a^2}{Q_r (Q_r + Q_a)^2 V_{eff}}, \quad (1)$$

where $E_{loc}$ and $E_i$ represent the localized field and the input field, respectively. $\kappa_i$ is the coupling coefficient with the external input field, which depends on the radiation channel and is expressed as $\kappa_i = \sqrt{2\gamma_r}$, where $\gamma_r = \omega/(2Q_r)$ denotes the radiative loss. $\omega$ and $Q_r$ represent the angular frequency and the radiative quality factor, respectively. In a lossy plasmonic structure, the total quality factor $Q_{tot}^{-1}$ is defined as $Q_{tot}^{-1} = Q_r^{-1} + Q_a^{-1}$, which combines the radiative loss $Q_r^{-1}$ and the non-radiative loss $Q_a^{-1}$. $V_{eff}$ denotes the normalized effective mode volume. For an undeformed lattice array, when the incident light wavelength matches the resonance wavelength of the BIC mode, the symmetry-protected BIC mode at normal incidence (Γ point) is decoupled from the plane wave resonance mode. Consequently, the ideal BIC cannot be excited by the incident light or coupled to the far field, as shown in the left of Fig. 1b. While it possesses perfect spatial purity due to its symmetry, this also results in a vanishing field enhancement factor ($G = 0$) and precludes any useful energy exchange or information flow. Trade-off Eq. (1), maximization of the field enhancement factor $G$ which converges to the critical state can be achieved by the radiative losses balancing the non-radiative losses ($Q_r = Q_a$)[50,51]. Therefore, $\sigma_h$ symmetry breaking was intentionally introduced into the aluminum nanopore array to convert quasi-BIC into finite Q BICs. This transition introduces controlled radiative leakage channels and enables stronger field confinement, resulting in a significant reduction of $V_{eff}$. Based on the undeformed flat crystal, we fabricated the TPC, as illustrated in the right panel of Fig. 1b. The three-dimensional morphology of the TPC was reconstructed as accurately as possible from AFM scanning images, with its thickness estimated to be approximately 220 nm based on the measured height difference (Supplementary Materials Fig. S3).

While the $\sigma_h$ symmetry was broken by hole etching slowly (Fig. 1c), the symmetric even modes of the columnar void would not be confined any longer when the size of the trunk is smaller than the cutting off size of the modes on columnar unit cell, which makes the even modes radiative [52–54]. Meanwhile, the transvers electric (TE) and transvers magnetic (TM) modes couple[55]. As the tip of the lattice became lower, the radiative angle was tuned (Fig. 1d). In addition, the fundamental $m = 0$ mode characterized by a vertically oriented electric dipole, does not exist in symmetric lattices owing to the constraint of mirror symmetry. The field concentration near the tip produces an ultra-small optical mode volume and substantial field enhancement through geometric focusing. Consequently, the $m = 0$ mode provides an efficient pathway for light–matter interaction enhancement and serves as a radiative channel complementary to the symmetry-protected q-BICs in the plasmonic cone array.

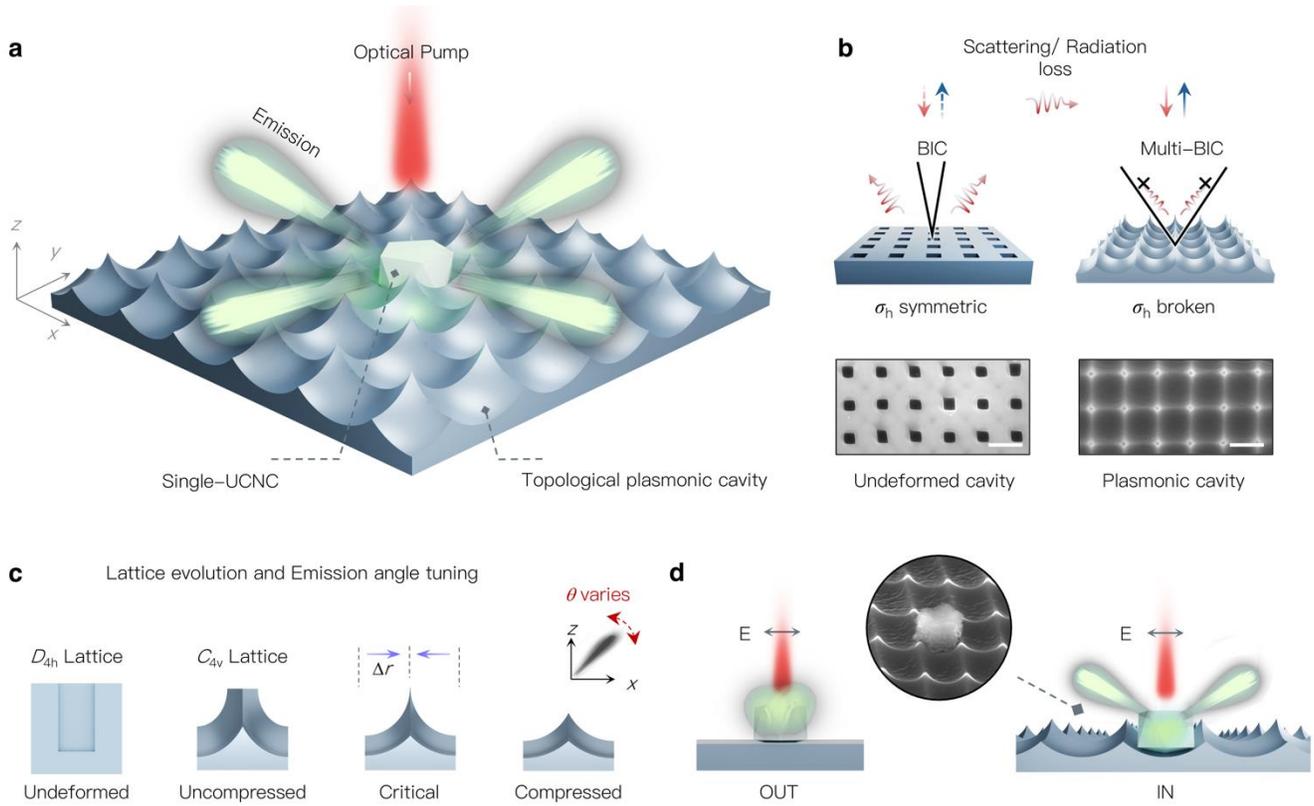

**Figure 1 Upconversion light field modulation of single-NC by $\sigma_h$ Symmetry-breaking TPC. a,** Schematic of the hybrid cavity-emitter system composed of a single-NC embedded in TPC. The periodic plasmonic nanostructures have a tetragonal lattice and form nanocones at the center of lattice. **b,** Left: In a $\sigma_h$ symmetric (undeformed) lattice, the symmetry-protected quasi-boundary excited state at the Γ point has extremely low radiation loss but weak coupling with free-space excited states, while also exhibiting limited emission field confinement. Right: Breaking $\sigma_h$ symmetry transforms the lattice into an asymmetric plasmonic cavity, which supports multi-BICs and exhibits enhanced field localization and large-angle field confinement. The bottom panel shows scanning electron microscope (SEM) images of undeformed and deformed plasmonic lattices, with a scale bar of 400 nm. **c,** the evolution of the lattice geometry from the undeformed state ($D_{4h}$) to the compressed state ($C_{4v}$) leads to a transition from q-BIC to multimode BIC. The tip of the compressed $C_{4v}$ lattice is used to achieve a tunable emission angle. The emission direction $\theta$ continuously evolves

with structural deformation, accompanied by the underlying band topology. **d,** Schematic illustrating directional control of upconversion emission. Without cavity coupling (OUT), emission is isotropic. In contrast, when embedded in the TPC (IN), the NC emission is funneled into well-defined lattice plasmonic mode cones, resulting in sharply confined directional radiation under normal incidence.

We highlight the evolution from symmetry-protected states to radiative states and the control mechanism for directional emission by comparing the evolution of the energy band structure in the plasmonic lattice across the breaking $\sigma_h$ symmetry. As shown in the top panel of Fig. 2a, the undeformed lattice of the two-dimensional square array belongs to the $D_{4h}$ symmetry group, exhibiting reflection symmetry $\sigma_h$ in the horizontal (*x-y*) plane. The corresponding TM-like polarized band structure is shown in Fig. 2a, where four bands are identified within the frequency range of interest. At the $\Gamma$ point, the three-dimensional mode distribution of the low-frequency band ($\Gamma_1$, 616 nm) exhibits a quadrupole mode (Fig. 2d), while the high-frequency band exhibits a dipole mode (see Supplementary Materials Fig. S1). Bands 2 and 3, under the $D_{4h}$ symmetry, form a degenerate pair characterized by distinct symmetry-protected modal distributions.

The undeformed lattice, after wet chemical etching, exposes the underlying barrier layer, breaking the $\sigma_h$ mirror reflection symmetry of the $D_{4h}$ point group (top panel of Fig. 2b) and transforming the undeformed lattice into a plasmonic lattices ($C_{4v}$) with a nanocone geometry. By tuning the compression parameter *r*, we force the bands to form accidental degeneracy near the $\Gamma$ point, and the corresponding band structure of the plasmonic cone-shaped lattice is shown in Fig. 2b. Under these ideal conditions (*r* = *a*/2), the perfect plasmonic lattice exhibits single-degenerate $\Gamma_1$ and $\Gamma_4$ modes at the $\Gamma$ point, which overlap with the doubly degenerate $\Gamma_{2/3}$ mode. This results in a fourfold degeneracy and a band gap closure, rendering the Zak phase ill-defined. To isolate the BIC mode, we further compressed the plasmonic nanocones, resulting in a lattice configuration with $\Delta r > a/2$, as shown in Fig. 2c. The corresponding band structure reveals a lifted degeneracy of the $\Gamma_4$ mode, where the frequency shift of these bulk guided modes effectively prepares the system for the deterministic control of radiation directionality. At the upper band $\Gamma$ point ($\Gamma_4$), the three-dimensional mode profile retains the characteristic quadrupolar field distribution of the undeformed unit cell (Fig. 2f). Although the $E_z$ component of the lower band mode is also a quadrupole BIC mode, it is submerged in band 2, 3. In contrast, for compressed nanocones with $\Delta r < a/2$ (Fig.1c), both the band structure and mode profiles clearly exhibit a Zak phase transition at the degeneracy point, confirming a band inversion (see Supplementary Materials Fig. S11).

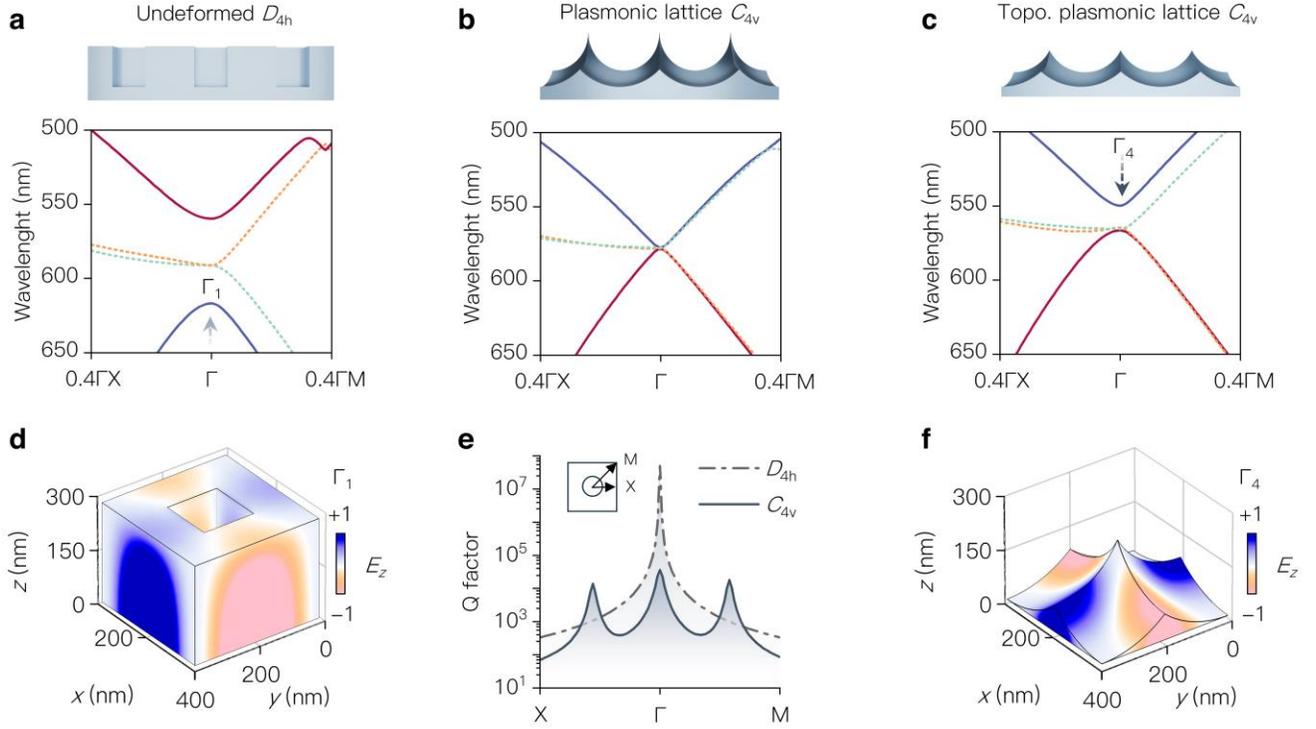

**Figure 2 Transition of $\sigma_h$ symmetry-breaking quasi-BIC to high Q multi-BIC. a**, Schematic and TM-polarized band structure of an undeformed square lattice with $D_{4h}$ symmetry, which preserves horizontal mirror symmetry ($\sigma_h$). Four photonic bands are observed near the frequency range of interest. At the Γ point, the lowest-frequency mode ($\Gamma_1$, 616 nm) exhibits a quadrupolar field distribution (shown in **d**). Bands 2 and 3 form a degenerate pair ($\Gamma_{23}$) protected by symmetry. **b**, In plasmonic lattice with cone-shaped geometry ($C_{4v}$ symmetry), $\sigma_h$ symmetry is broken via guided anodization and wet etching to reach a critical state ($\Delta r=a/2$). The resulting band structure features accidental degeneracy at the Γ point between $\Gamma_1$, $\Gamma_4$, and $\Gamma_{23}$ modes. **c**, Further compression of the plasmonic nanocones ($\Delta r > a/2$) lifts the degeneracy of $\Gamma_4$, resulting in a topological plasmonic cavity with finite mode volume and well-separated bands. The upper $\Gamma_4$ mode preserves quadrupole symmetry, as visualized in **f**. **d**, Three-dimensional field distribution ($E_z$ component) of the $\Gamma_1$ mode in the undeformed lattice. **e**, Q-factor evolution of BIC modes for both undeformed ($D_{4h}$) and $\sigma_h$ symmetry-broken ($C_{4v}$) plasmonic lattices. The BIC in the $D_{4h}$ lattice exhibits an infinite Q-factor, while symmetry breaking induces finite Q multi-BICs (~$10^4$). **f**, Three-dimensional $E_z$ field profile of the $\Gamma_4$ mode in the compressed topological plasmonic cavity, retaining a quadrupolar character but now with controlled radiative leakage.

This behavior indicates that, in addition to the conventional dynamical phase, the crystal band also acquires a topologically nontrivial geometric phase during the optimization of the sharp cone. For the quadrupole bands in both lattices, the band at the Brillouin zone center of the undeformed lattice exhibits a theoretical BIC with an infinite Q-factor, as indicated by the dashed line in Fig. 2e. In the topological plasmonic lattice ($\Delta r > a/2$), breaking the $\sigma_h$ mirror symmetry transforms the quasi-BIC into a multi-BIC (the three peaks in Fig. 2e and black arrows indicated in Fig. 3e) with a finite Q-factor (~$10^4$), thereby enabling controlled energy leakage to external plane waves, as shown by the solid line in Fig. 2e. Notably, the topological plasmonic lattice supports multi-BIC (*48*), which, compared to the quasi-BIC, exhibit a broader confinement angle in momentum space. This feature further enriches the system's directional emission characteristics.

To experimentally characterize the dispersion properties of the plasmonic lattices with broken **$\sigma_h$** mirror symmetry, we have established an angle-resolved scattering spectroscopy (ARSS) setup based on a 4*f* optical system capable of rapidly switching between real and momentum space (Supplementary Materials Fig. S8). We measured the ARSS of the prepared undeformed lattice cavity, plasmonic cavity, and topological plasmonic cavity, and compared them with the angle-resolved reflection spectra (ARRS) from rigorous numerical simulations based on finite-element method.

As shown in Fig. 3, the experimental ARSS (a–c) and simulated ARRS (d–f) provide a comprehensive map of the lattice evolution and its impact on radiation. For the undeformed cavity (Figs. 3a, d), the spectra confirm the presence of a symmetry-protected BIC at the $\Gamma$ point. Upon breaking the $\sigma_h$ mirror symmetry (Figs. 3b, e, $\Delta r = a/2$), the emergence of localized surface plasmon resonances (LSPR, red arrows) and multi-BIC modes (black arrows) serves as a clear signature of the deformed nanocone geometry. While these features contribute to field enhancement, the deterministic control of emission directionality is primarily driven by the strong hybridization between TE and TM modes.

Crucially, the introduced $\sigma_h$ symmetry breaking lifts the modal orthogonality, inducing a TE and TM modes hybridization that manifests as a distinct flat band intersecting the TM dispersion branches. This hybridization is clearly observed in both experimental and simulated spectra (Figs. 3b, e), becoming more pronounced in the fully compressed topological cavity (Figs. 3c and f, $\Delta r > a/2$). This mechanism effectively opens a specific radiation channel at 550 nm; because the hybridized band is remarkably flat, it possesses a near-zero group velocity, facilitating the efficient leakage of trapped light into a deterministic far-field direction. The high degree of consistency between the experimental ARSS and simulated ARRS confirms that this mode-coupling-induced leakage channel establishes the requisite physical foundation for the directional enhancement of upconversion emission from single quantum emitters.

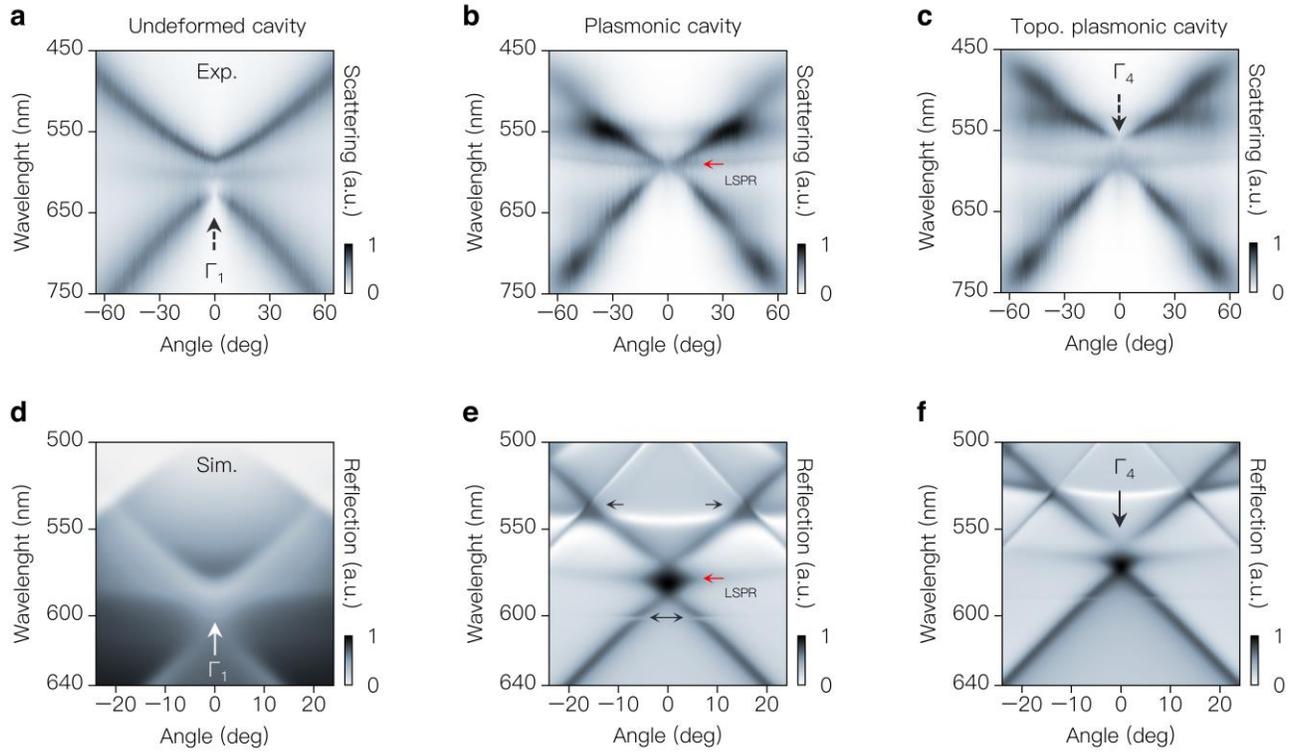

**Figure 3 Angle-resolved spectroscopy reveals topological phase transition and the emergence of LSPR after $\sigma_h$ symmetry breaking. a–c,** Experimental ARSS of the (a) undeformed cavity, (b) plasmonic cavity, and (c) topological plasmonic cavity. The undeformed $D_{4h}$-symmetric cavity (a) supports a quasi-BIC at the Γ point in the lower-frequency $\Gamma_1$ band, exhibiting vanishing reflectance due to its decoupling from the far field. Upon $\sigma_h$ symmetry breaking (b), the lattice evolves into a $C_{4v}$ symmetric plasmonic cavity, leading to an accidental fourfold degeneracy at the Γ point ($\Delta r = a/2$), characterized by band touching and topological phase ill-definition. Further compression of the nanocones (c, $\Delta r > a/2$) lifts this degeneracy, resulting in the emergence of a finite Q BIC ($\Gamma_4$) in the upper band near 550 nm. The appearance of this resonance marks a topological transition accompanied by a band inversion. **d–f,** Corresponding simulated ARRS of the three cavities, compared with simulated angle-resolved resonance spectra. The $\Gamma_1$ mode in the undeformed cavity (d) remains dark at the Γ point, while in the plasmonic cavity (e), a distinct localized surface plasmon resonance state emerges near the flat band (red arrow), arising from the extremely compressed cone tips. This LSPR persists in the topological plasmonic cavity (f), alongside the finite Q multi-BIC mode ($\Gamma_4$ black arrow).

## Far- and near-field characterizations

To elucidate the near-field enhancement and directional emission characteristics of BIC modes, we conducted both near-field and far-field analyses for the undeformed and topological plasmonic cavities. Figs. 4a and 4b show the simulated *xz*-plane electric field profiles ($E_z$) at the BIC resonance wavelengths for the conventional BIC ($\Gamma_1$) and the finite Q q-BIC mode ($\Gamma_4$), respectively. The undeformed cavity exhibits vertical symmetry and a flat-top surface, with its near-field enhancement effect confined within the cavity. In contrast, the topological plasmonic cavity exhibits significant electric field localization (highest LDOS) at the nanocone tip, this is attributed to the extreme mode compression caused by the geometric conical structure, which enhances the surface electric field intensity through strong field hotspots. Additionally, the angular distribution of the induced scattered light under normal incidence conditions is redistributed, as indicated by the tilted field lobes in Fig. 4b.

To further investigate the spatial extent of field localization, we simulated the normalized electric field distributions in the *xy*-plane over a 6 × 6 supercell (Figs. 4c and 4d). In the undeformed cavity ($D_{4h}$ symmetry, Fig. 4c), the near field extends into the surrounding air, indicating weak lateral confinement. In contrast, the topological plasmonic lattice ($C_{4v}$ symmetry, Fig. 4d) demonstrates strong in-plane field confinement, primarily due to the formation of multiple finite Q BIC modes within the unit cell. This effective confinement forms the basis for enhanced light–matter interaction in the system.

We next evaluated how these two cavity types modulate the far-field radiation pattern of an embedded single-NC, modeled as a point electric dipole at the center of the supercell. As shown in Figs. 4e and 4f, the undeformed cavity directs the emission into three main beams, including a prominent vertical lobe, whereas the topological cavity suppresses vertical emission and promotes dual-beam radiation at oblique angles (~56°). This directional pattern is a direct manifestation of the symmetry-broken quasi-BIC nature of the $\Gamma_4$ mode, where vertical dipole components are suppressed and lateral radiation dominates.

The deterministic steering of upconversion emission through structural deformation is demonstrated in Figs. 4g and 4h. In the uncompressed nanocone lattice ($\Delta r = 0.4a$, Fig. 4g), a single-NC radiates two closely spaced emission beams under x-polarized excitation. However, as the lattice is further compressed to $\Delta r = 0.55a$ (Fig. 4h), these beams undergo significant angular separation, resulting in highly directional emission. This evolution confirms that the **$\sigma_h$** symmetry breaking, combined with the deliberate lattice design, induces strong TE and TM mode coupling, providing a robust physical mechanism for tailoring far-field radiation. Fig. 4i quantitatively summarizes this angular tunability by superimposing emission curves across various nanocone radii, demonstrating that this mode-hybridization-driven approach enables precise control over the emission angle from a single point source.

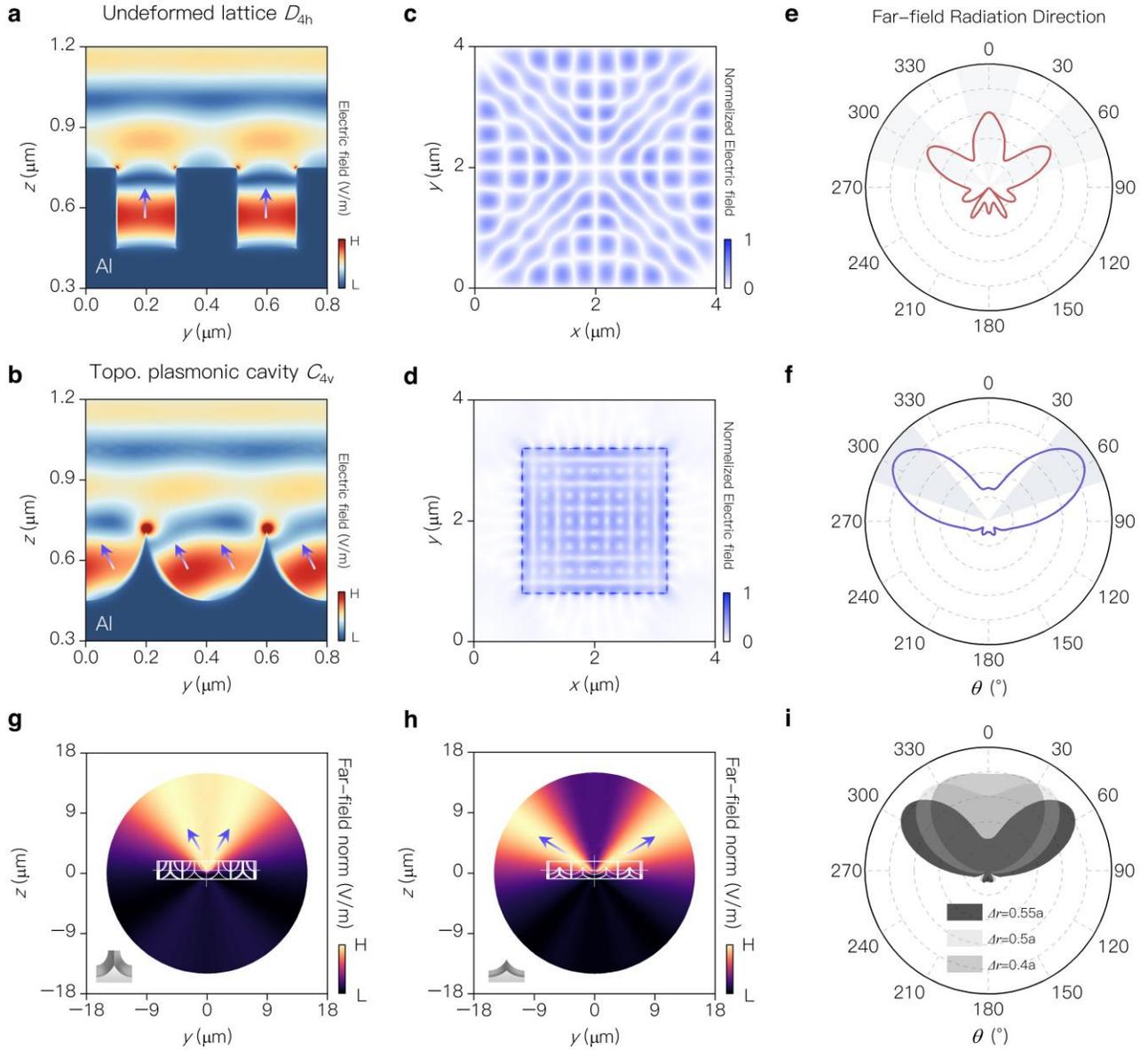

**Figure 4 Near-field enhancement and directional emission of undeformed and topological plasmonic cavities. a,b,** Simulated *xz*-plane electric field profiles at the BIC resonances of the undeformed ($\Gamma_1$) and topological plasmonic ($\Gamma_4$) cavities. The blue arrow emphasizes the near-field Poynting vector. **c,d,** *xy*-plane near-field maps over a 6×6 supercell reveal improved in-plane confinement in the topological cavity due to multi-BIC formation. **e,f,** Far-field emission profiles for a single upconversion dipole placed at the supercell center: the topological cavity suppresses vertical emission and supports two symmetric large-angle beams. **g,h,** Far-field distributions under *x*-polarized excitation for cone base radii $\Delta r = 0.4a$ and $\Delta r = 0.55a$ ; increasing $\Delta r$ leads to beam divergence. **i,** Polar plot of emission angle $\theta$ for different cone radii, demonstrating emission angle control via nanocone tapering.

**Experimental demonstration**

To verify the robustness of the topological plasmonic cavity against local perturbations, we characterized the system's response upon the introduction of a single upconversion nanoparticle (UCNP, ~400 nm diameter) into the lattice. For comparison, the emission of an isolated UCNP on a flat aluminum film characterized by near-isotropic radiation and broad angular distribution is detailed in the Supplementary Information. Fig. 5a presents the simulated and measured mode distributions of the cavity before and after the placement of the nanoparticle. Despite the presence of the sub-wavelength scatter, the BIC mode remains spatially stable and structurally intact. This visual consistency confirms that the topological nature of the cavity suppresses scattering loss into the continuum, maintaining a high-quality factor even with local symmetry breaking.

This robustness is further quantitatively validated by the experimental angle-resolved scattering spectra shown in Fig. 5b. The angular dispersion remains remarkably clean and well-defined after the nanoparticle loading, with the characteristic resonance features showing negligible broadening or shift. The fact that the cavity's momentum-space signature is preserved against small-particle scattering highlights the practical advantage of using topologically protected states for stable, directional light-matter interactions.

However, in the absence of structural modulation or symmetry-defined mode selection, free-space NC emission remains omnidirectional, resulting in poor far-field coupling and limited suitability for directional beam shaping. To overcome this limitation and to explore cavity-enhanced emission dynamics, we deposited a dilute dispersion of NCs directly onto the topological plasmonic cavities. Due to the comparable size between the NC particles and the lattice pitch, isolated NC preferentially settle into the interstitial sites of the lattice, forming well-defined NC–cavity hybrid systems. One such NC particle was found to be precisely positioned at the center of a four-unit supercell gap (see Supplementary Materials Fig. S6).

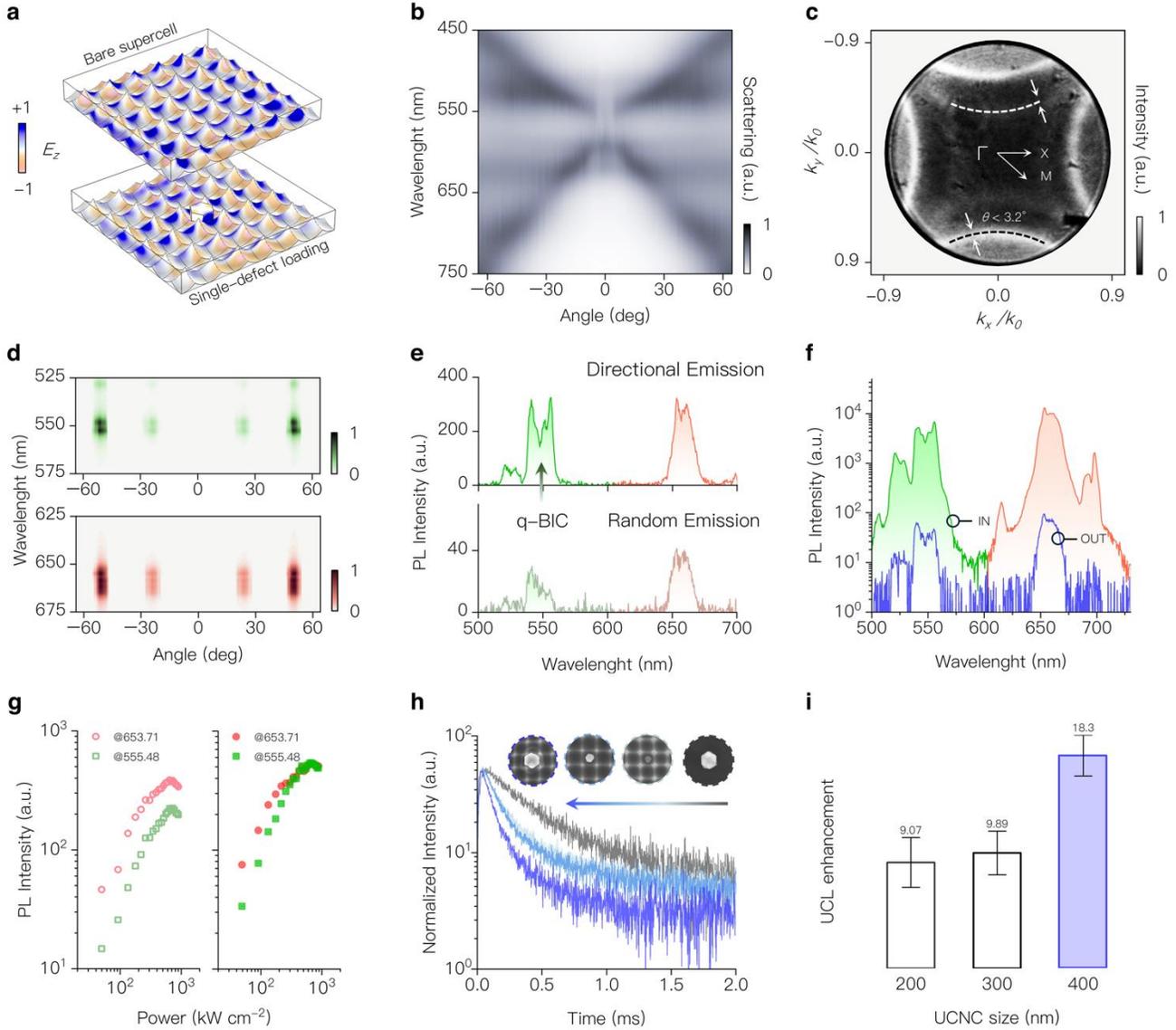

**Figure 5 Directional beam shaping and Purcell-enhanced upconversion emission from a single-NC coupled to TPC.**
**a**. Simulation of the real-space mode distribution of a topological plasmonic cavity before and after loading a single 400 nm upconversion nanoparticle (UCNP). **b**. Experimental ARSs of the cavity loaded with a single UCNP. **c**. Fourier-plane image of a single NC coupled to TPC, revealing a highly directional crescent-shaped emission cone confined within ±3.2° (black dashed arc), indicating strong coupling between the NC and the multi-BIC of the TPC. A weaker secondary emission cone is also observed at ±26° (white dashed arc). **d**, Angle-resolved photoluminescence spectrum of the single-NC–TPC hybrid along the X–Γ–X direction. The outer ring (±56°) and inner ring (±26°) exhibit narrow-band strong emission aligned with the open radiation channel between multi-BICs. **e**, Comparison of emission spectra between a NC outside (bottom) and inside (top) the TPC. The 550 nm emission ($^4S_{3/2} \rightarrow {}^4I_{15/2}$) is significantly enhanced due to spectral overlap with the high-Q cavity mode. **f**, Corresponding upconversion spectra in the TPC region (green and red curves) and outside the TPC region (blue colour curves). **g**, Power-dependent PL spectra showing emission enhancement dynamics. With increasing 980 nm excitation power, the 550 nm emission overtakes the 650 nm transition, indicating cavity-induced stimulated emission via the q-BIC mode. **h**, Time-resolved PL decay curves of single-NCs with different diameters (200 nm, 300 nm, 400 nm) deposited on the TPC. The 400 nm NC shows the greatest lifetime reduction, indicating strong radiative rate enhancement via Purcell effect. **i**, Extracted Purcell factors for NCs of different sizes. A maximum Purcell factor of ~17 is achieved for 400 nm NCs.

Under 980 nm excitation, the upconverted PL from the TPC and single-NC hybrid was projected to the Fourier back focal plane (Fig. 5c). The NC emission, which originally displayed near-isotropic radiation on a flat aluminum film (detailed in Supplementary Note 8), is now reshaped into a sharp, crescent-shaped emission cone in momentum space (outlined by the black dashed arc). The measured angular divergence is below 3.2°, representing a dramatic enhancement in directionality. This behavior arises from the resonant coupling between the NC emission and the surface plasmon polariton (SPP) band of the topological cavity, demonstrating the robust beam-shaping capability of the BIC-based system even under single-defect loading. The periodic plasmonic potential reshapes the wavevector of the upconverted light, effectively folding it into Bloch modes governed by the cavity symmetry. Notably, the crescent deformation near M point reflects a gradual breakdown of the designed momentum constraint, where coupling to off-symmetry SPP modes becomes less efficient (*56*). Thus, the engineered lattice modulates both the spatial and angular distribution of a single emitter, enabling well-defined directional upconversion at the nanoscale. The appearance of this sharply confined momentum-space feature confirms the efficient coupling between the NC emission and the designed plasmonic multi-BIC states.

Angle-resolved photoluminescence measurements were performed on a single-NC coupled to a topological plasmonic cavity, and momentum space profiles were obtained by scanning along the highly symmetric X-Γ-X direction (Fig. 5d). The outer emission ring corresponding to ±56° appears particularly bright and spectrally narrow, marking the polarization exciton mode emission cone (black dashed arc in Fig. 5c). Its characteristics in momentum space are highly similar to those of finite Q q-BIC in the cavity band structure (see Supplementary Materials Fig. S10), indicating that upconversion emission can be strongly influenced by the underlying plasmonic dispersion. Additionally, a weaker inner ring centered at ±26° also appears (the white dashed arc in Fig. 5c), which is not clearly visible in the original Fourier plane image. Spectrally, the resonantly enhanced emission under cavity coupling is evident in Fig. 5e. Compared to the randomly emitting NC outside the cavity (bottom panel), a single-NC inside the TPC shows significant enhancement of the 550 nm emission (top panel), corresponding to the $Er^{3+}$ transition that overlaps with the q-BIC cone. This directed emission is uncommon in the absence of a cavity.

To investigate the enhancement of exciton emission via $\sigma_h$ symmetry-breaking nanocavity, upconversion emission from single-NC was measured directionally at the resonance Fig. 5f compares the upconversion emission signals of a single-NC placed outside (blue curve) and inside (red curve) the TPC. Due to the small $V_{eff}$ of LSPR supported by $\sigma_h$ symmetry breaking TPC and the high Q factor characteristics of multiple BICs, the upconversion emission intensity of single-NC in TPC is enhanced by two orders of magnitude. This enhancement becomes more apparent under power-dependent excitation. As shown in Fig. 5g, with increasing pump intensity, the 550 nm emission gradually increases and eventually surpasses the 650 nm off-resonant transition, indicating stimulated emission gain mediated by the q-BIC band. This crossover marks a transition from spontaneous to directional stimulated upconversion through BIC-assisted mode selection. To evaluate the Purcell enhancement effect of upconversion emission from single-NCs in the cavity, we performed time-resolved fluorescence spectroscopy measurements using femtosecond laser excitation (Supplementary note3). UCNCs of different diameters (200 nm, 300 nm, and 400 nm) were sparsely spin-coated on the TPC, and the decay curves of upconversion emission from single-NCs were measured (Fig. 5i). Among them, the 400 nm NCs showed the most significant lifetime shortening in the cavity compared to the uncoupled samples, confirming the strong radiative rate enhancement. The Purcell factors extracted in each case are compared in Fig. 5i, with the 400 nm NCs showing up to ~17-fold enhancement. The corresponding UCL decay kinetics and analysis details are provided in the Supplementary Note5.

**Conclusion**

In summary, we have experimentally demonstrated significantly enhanced directional upconversion emission from a single nanocrystal (NC) enabled by a $\sigma_h$ symmetry-broken topological plasmonic cavity (TPC) platform. We developed a mask-free, high-throughput, and low-cost fabrication strategy based on guided anodic oxidation and wet chemical etching to realize TPC architectures with explicitly broken $\sigma_h$ mirror symmetry. By breaking the vertical mirror symmetry, we drive the transition of the lower-branch lattice plasmonic mode from symmetry-protected bound states in the continuum (BICs) to high-quality-factor multi-BIC regimes at the Γ point, which generates intense localized surface plasmon resonance (LSPR) hotspots and simultaneously opens controllable far-field radiation channels. The plasmonic lattice is judiciously engineered in momentum space such that its quasi-BIC (q-BIC) resonance spectrally overlaps with the characteristic 550 nm upconversion emission of dopant ions, thereby enabling bright, directional, and strongly amplified upconversion radiation from a single NC.

This work explicitly unveils the physical mechanism underlying q-BIC-enhanced upconversion luminescence and directional far-field emission of single nanocrystals in periodic plasmonic lattices, establishing a low-cost, high-efficiency nanoscale coherent light source with locked and controllable wide-angle emission directionality. Benefiting from the strong coupling between q-BIC modes and single emitters in symmetry-broken TPCs, the proposed platform exhibits outstanding structural robustness against local perturbations and overcomes the inherent limitations of conventional ensemble-based emitter systems. Our strategy offers a universal and scalable route for engineering deterministic directional light emission at the single-particle level, paving the way for advanced applications including high-resolution 3D projection imaging, on-chip directional optical antennas, single-photon nonlinear devices, and coherent imaging systems integrated with augmented reality waveguides.

## Methods

**Characterization.** The surface characteristics of the samples were captured using a scanning electron microscope (SEM, MIRA3-LMH) operating in in-beam secondary electron mode at an accelerating voltage of 20 *kV*. For atomic force microscopy (AFM) imaging, a material-based AFM (BRUKER, Dimension FastScan) was used in contact mode, with a scan rate set for high-resolution imaging. The AFM images were processed and analyzed using NanoScope Analysis 3.00 software, which produced a detailed 3D topographical profile of 2 *μm* × 2 *μm*.

**Optical characterization (ARSS, ARPL and UCL lifetime).** Both ARSS and ARPL measurements were performed using a Fourier back-plane (FBP) imaging system (supporting). For ARSS experiments, a halogen lamp was used as the excitation source. The polarization direction of the incident light was controlled by a rotatable adjustable mount with a linear polarizer, which aligned the polarization along the *y*-direction before the light entered a beam-splitting prism. A dark-field module (model) was employed to direct the incident light onto a 100× Olympus objective lens, which had a high numerical aperture (NA = 0.9) and an infinite conjugate design. The light was incident at a steep angle to the sample surface. The diffracted light was then collected by the same objective lens and sent into a 4f optical system composed of three lenses and a pinhole. The Fourier plane of the diffracted light was projected onto a CCD camera (QImaging Retiga R1 from Cairn Research) and a fiber spectrometer (Exemplar®Plus, BTC655N) through beam splitter BS2. Spectral signals were gathered using a fiber with a core diameter of 100 μm (model), which was positioned to collect light spots on the Fourier plane. The fiber's collection end was mounted on a support rod connected to an electric displacement platform, allowing for precise point-by-point scanning with a 0.1 mm step. This setup enabled efficient dispersion imaging in momentum space and minimized stray light interference, ensuring sharp resonance peaks. The ARPL measurements were performed using the same optical configuration, i.e. a single wavelength continuous laser ($\lambda = 980$ nm) was used to vertically excite single-NC. The spotlight signal from the sample was passed through an 800 nm short-pass filter, capturing both the Fourier image and the ARPL spectrum using the FBP system.

**Numerical Modeling.** The dispersion properties of the TPC structure and the emission enhancement of single-NCs in its lattice array were modeled using the commercially available software COMSOL Multiphysics. The three-dimensional model of the plasmonic lattices was obtained by Boolean operations on a tetragonal prism and an ellipsoid (long axis $a$, short axis $b$), where the ellipsoid's center is highly aligned with the surface of the tetragonal prism. The three-dimensional energy band structure of the plasmonic lattices and the Q-factors of the corresponding bands were calculated using eigenfrequency studies with periodic boundary conditions, and redundant pseudo-modes were manually eliminated. Three-dimensional mode analysis and angle-resolved reflectance spectra of this periodic structure were calculated using frequency-domain studies.

**Supporting information**

TPC fabrication, Synthesis of NaYF$_4$:10% Yb$^{3+}$, 2% Er$^{3+}$@NaYF$_4$ core-shell nanocrystals, undeformed and plasmonic lattices Γ-point mode, Fourier back plane image of TPC, Comparison of $D_{4h}$ and $C_{4v}$ cavities for near-field enhancement are included in the supporting information at https://xxx.xxx.

**Acknowledgements**

The authors thank the supporting of the National Key Research and Development Program of China (2024YFA1409900), National Natural Science Foundation of China (Grant No. 12274054, U24A2018).

**Author contribution**

Y.F. conceived the idea and directed the project. Y.C. performed the main experiment and the main FEM calculations. M.Z performed the cavity substrate fabrication. Q.B. performed the time resolved florescence.

X.Y. performed the upconversion nanoparticle synthesis. B.D. directed the upconversion nanoparticle synthesis. W.W directed the cavity fabrication. Y.F. and Y.C. discussed and concluded the main results. Y.C. wrote the manuscript. All of the authors revised the paper.

**Conflicts of interest**

The authors declare no competing financial interest.

# Supplementary Materials for

# Robust topological BIC nanocavities for upconversion directional emission

Yongqi Chen[1], Ming Zhu[2], Qingfeng Bian[1], Xiumei Yin[3], Wenxin Wang[2, *], Bin Dong[3,*], and Yurui Fang[1, *]

[1.] School of Physics, Dalian University of Technology, Dalian 116024, China.
[2.] Qingdao lnnovation and Development Center, Harbin Engineering University, Qingdao, 266500, China
[3.] School of Physics and Materials Engineering, Dalian Nationalities University, Dalian 116600, China
*Corresponding authors: wenxin.wang@hrbeu.edu.cn (W.W), dong@(B.D.), yrfang@dlut.edu.cn (Y.F.)

## Contents

# 1. Undeformed and TPC lattice Γ point mode profiled

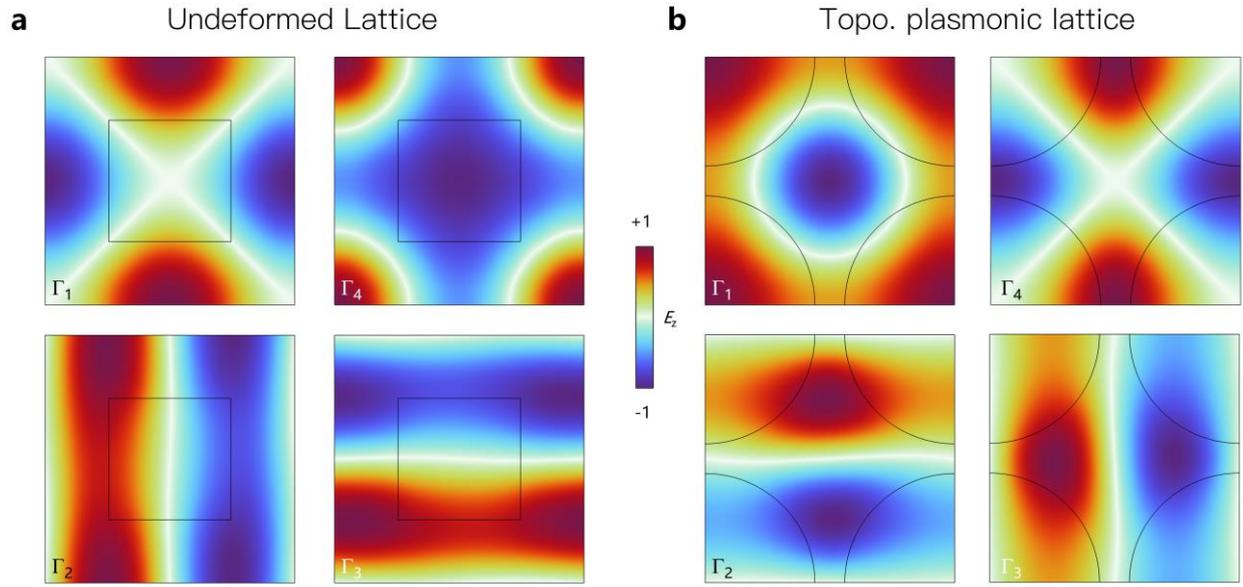

**Supplementary Fig. S1 Shows the mode distributions at the Γ point for undeformed (a) and plasmonic lattices (b) under periodic boundary conditions.** To reveal the symmetry-protected BIC, we calculated the eigenfrequencies of both lattices located at the Γ point. These mode distributions illustrate the normalised $z$ component of the electric field $E_z$ in the lattice $xy$ plane. **a,** In aluminium photonic crystal plates, the undeformed lattice (lattice constant $a = 400$ nm, air-hole edge length $r = 0.49a$, height $H = 300$ nm) exhibits two BIC modes $\Gamma_1/\Gamma_4$, decoupled under $C_2$ symmetry. **b,** For the uncompressed plasmonic lattice ($\Delta r = 0.45a$), the quadrupole mode $\Gamma_1$ appears in the low-frequency branch, two coupled dipole modes in $\Gamma_2/\Gamma_3$, and a monopole mode ($\Gamma_4$) at the Γ point in the upper band.

## 2. Experimental samples

### 2.1 Undeformed cavity and plasmonic cavity fabrication

High-purity aluminum foil is cleaned using a solution of acetone and ethanol to remove surface impurities and oxides. The aluminum foil is then electropolished to create a smooth surface, providing a uniform base for the subsequent anodization process. A two-step anodisation method is used to prepare the tetragonal lattice aluminium photonic crystal hole plate shown in Supplementary Fig. S2a. In the first step, anodisation is performed under appropriate voltage conditions to form an anodised aluminium oxide (AAO) template layer on the aluminium photonic crystal plate. Subsequently, the initial AAO layer is completely dissolved in a chromic acid solution to expose a smooth and structurally regular aluminium surface.

In a second anodizing step, the aluminum surface is further anodized under the same voltage conditions, resulting in the formation of a highly ordered tetragonal AAO pore lattice(*1*). By adjusting the anodization voltage, duration, and electrolyte composition, the pore size, wall thickness, and overall thickness of the AAO template can be controlled. To form the periodic barrier layer at the bottom of the triangular lattice AAO template, anodization is continued until the pore growth stabilizes. Finally, the top oxide layer was removed by wet chemical etching to expose the plasmonic lattice array on the aluminium substrate (Supplementary Fig. S2b).

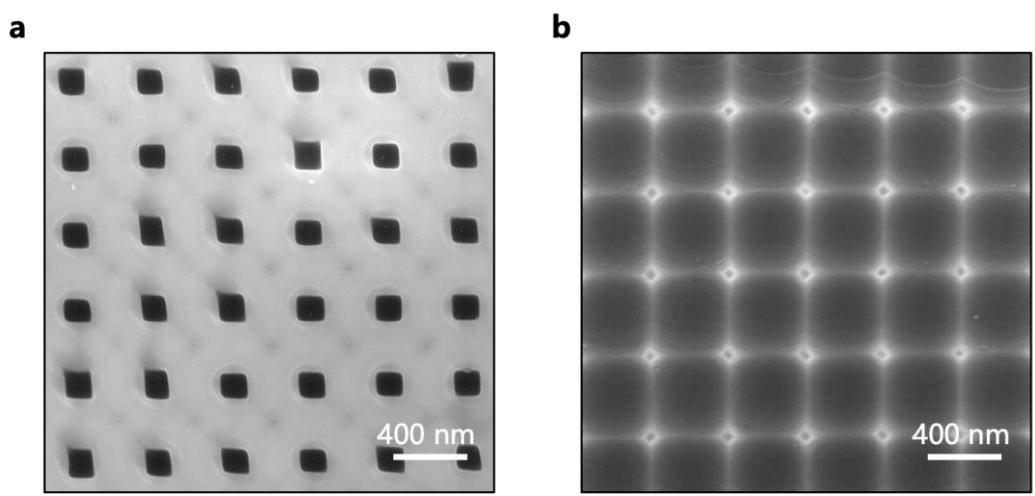

**Supplementary Fig. S2** The corresponding SEM image of undeformed aluminium photonic crystal hole plates **(a)** and plasmonic cavity lattice arrays **(b)** with lattice constant $a = 400$ nm from a perspective view.

## 2.1.1 Atomic force microscopy (AFM) image of the topological plasmonic cavity.

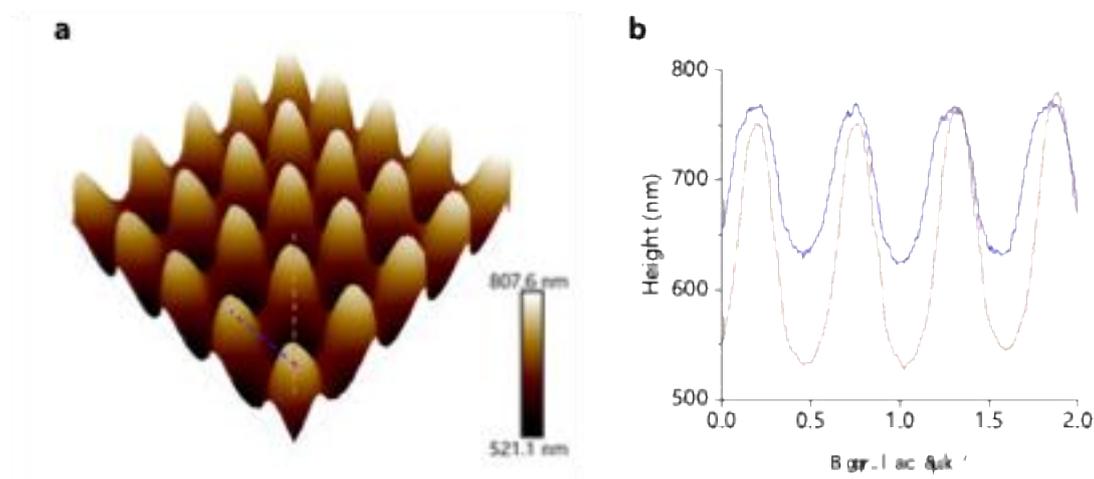

**Supplementary Fig. S3. a,** AFM three-dimensional image of the topological plasmonic cavity. **b,** The solid orange line shows the *XZ* height profile scanned along the diagonal path of the grid (the orange dashed line in (a)), revealing a height difference of 210 nm. The blue solid line shows the *XZ* height profile scanned along the direction of the right-angle side (blue dashed line in (a)).

## 2.2 Upconversion NCs sample information

### 2.2.1 Synthesis of NaYF$_4$:10% Yb$^{3+}$, 2% Er$^{3+}$@NaYF$_4$ core-shell NCs

3 mL oleic acid and 7 mL octadecene were added to the two vials respectively, and 2 mL Re(CH$_3$COO)$_3$ (0.2 M) was added in proportion (where Re = Y$^{3+}$, Yb$^{3+}$, and Er$^{3+}$ in a molar ratio of 88:10:2). The samples present pale yellow after heating to 150 °C and preservation for 45 min, a rare earth oleate precursor solution can be obtained while the sample was cooled to room temperature. 1 mL NaOH (1M) and 4 mL NH$_4$F (0.4 M) methanol mixture were added to the solution, insulated at 50 °C for 40 min. Vacuumed at 100 °C and argon was swap every 3 min to balance bottle pressure. The solution was quick heated (15 min) to 290 °C when bubble no longer produced, kept warm for 1.5 h, and desired product can obtain while cooled to room temperature. The nanoparticles with different size were synthesized at different temperature. A certain amount of ethanol solution was added to the solution, centrifuged at 9000 rad/min for 6 min. the product was dissolved in 4 mL cyclohexane, 8 mL ethanol was added and centrifuged at 9000 rad/min for 6 min. Repeat the previous step with 4 mL ethanol and 4 ml methanol instead of 8 mL ethanol. The resulting sample was dissolved in 5 mL cyclohexane sealed and stored in a glass vial at 4 °C.

Shell precursor solution was prepared as the preparation method of core layer and cooled to room temperature. As-prepared nuclear solution was injected into precursor, NaOH and NH4F mixed solution were added and incubated at 50 °C for 40 min. The sample was heated to 290 °C incubated for 2 h after the vacuum process. Original washing step was repeated to obtain the desired product. Dissolved in 5 mL of cyclohexane to be tested.

**2.2.2 Characterization of the as-produced NCs**

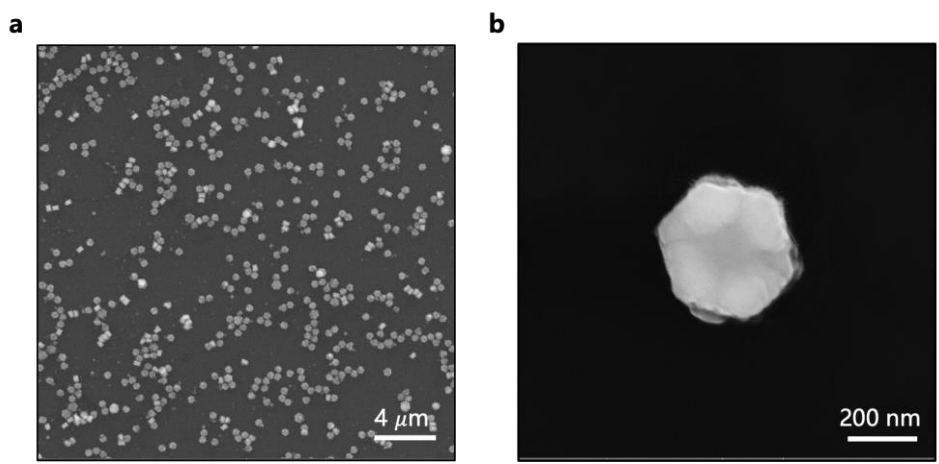

**Supplementary Fig. S4 SEM images of the synthesised core-shell NCs**

**2.2.3 Energy-level diagram and the multistep upconverison process for a NC**

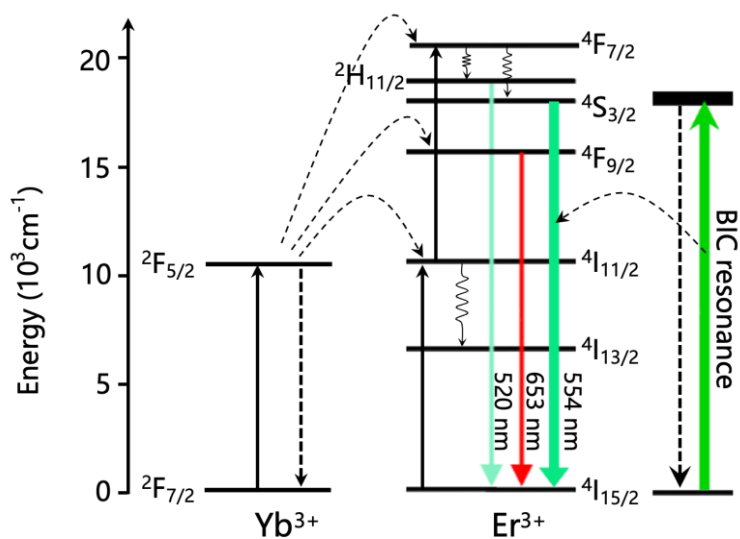

**Supplementary Fig. S5. Energy level of $Yb^{3+}$ and $Er^{3+}$ ions relevant to the energy-transfer upconversion process.**

The emission for luminescence up-conversion in $Yb^{3+}$ and $Er^{3+}$ co-doped $NaYF_4$ has been studied previously(*2–4*). The major processes are shown in Figure S5. Due to the large absorption cross-section of $Yb^{3+}$ and its higher doping concentration, most of the incident light is absorbed by $Yb^{3+}$, exciting it from the $^2F_{7/2}$ to $^2F_{5/2}$ state. Through an energy transfer process, the $Yb^{3+}$ ion excites a nearby $Er^{3+}$ ion into the $^4I_{11/2}$ state. Before returning to the ground state, another energy transfer occurs, which excites the $Er^{3+}$ ion to the $^4F_{7/2}$ level. Then, a non-radiative transition to the $^2H_{11/2}$ and $^4S_{3/2}$ states, the ions return to the ground state $^4I_{15/2}$, emitting green luminescence. The multi-BIC resonance supported by the TPC exhibits spectral overlap with the $^4S_{3/2}$ excited state emission of $Er^{3+}$-doped upconversion nanoparticles. A fraction of $Er^{3+}$ ions in the $^4S_{3/2}$ state would decay non-radiatively into the slightly lower $^4F_{9/2}$ level, followed by a transition back to the ground state, where red luminescence originates.

**2.3 Deposition of single-NCs in topological plasmonic cavities**

A 5 μL suspension of 400 nm NCs (5 mg/mL) was spin-coated onto the topological plasmonic cavity at a target speed of 4000 rpm. The spin-coating duration was 60 *s*, with the rotation speed ramping up to 4000 rpm within the first 10 s. We successfully placed an NC at the centre of the gap surrounded by four topological lattices. The SEM image is shown in Supplementary Fig. S6.

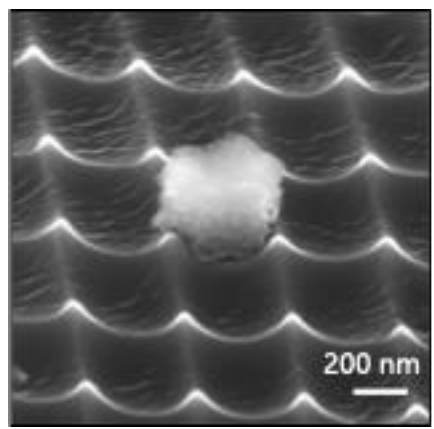

**Supplementary Fig. S6** A cavity-emitter system formed by the coupling of a single-NC with a particle size of 400 nm and a topological plasmonic cavity.

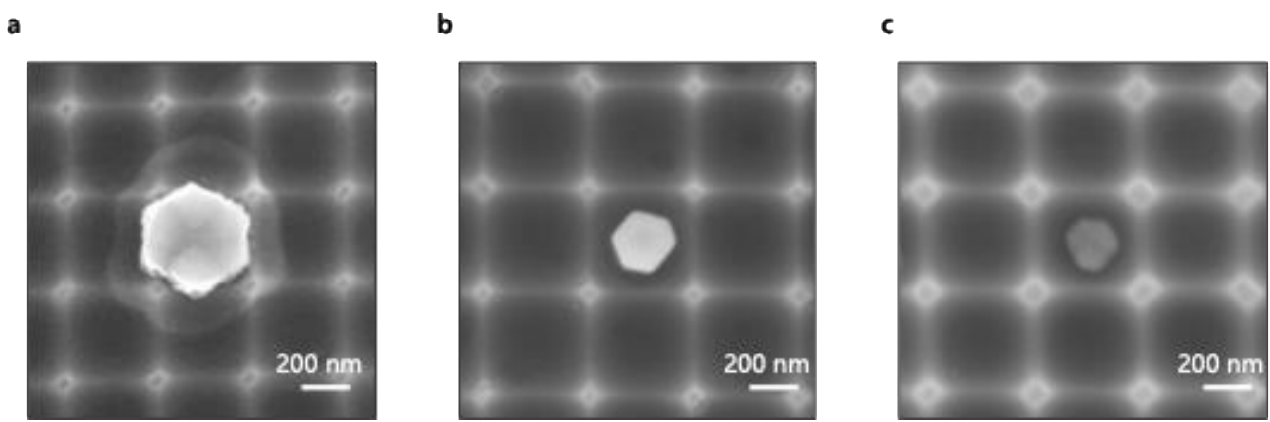

**Supplementary Fig. S7** Using the same spin-coating protocol, single-NCs of different sizes were readily coupled into the topological plasmonic cavity: **a,** 400 nm NC. **b,** 300 nm NC (5 mg/mL) at 4000 rpm, and **c,** 200 nm NC (3 mg/mL) at 4000 rpm.

## 3. Experimental setup

### 3.1 Common optical path for angle-resolved resonance scattering and photoluminescence

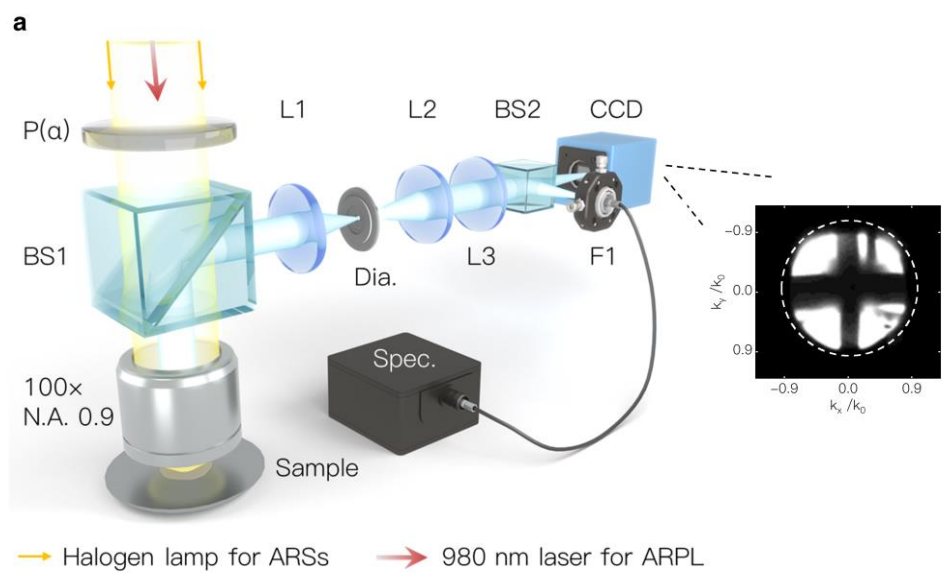

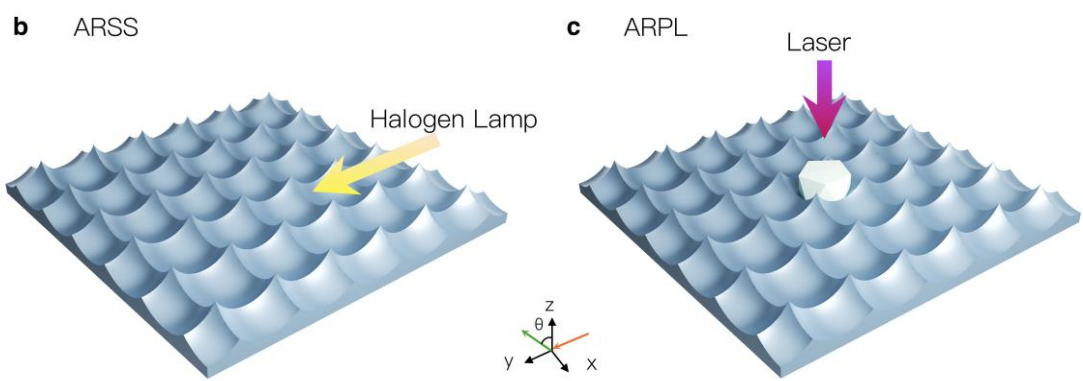

**Supplementary Fig. S8. Detailed optical setup for angle-resolved scattering spectroscopy and photoluminescence spectroscopy measurements.**

ARSS and ARPL were both based on a FBP imaging device. For ARSS measurements, the sample was excited using a halogen lamp as the incident light source. A rotatable adjustable mount with a linear polarizer was used to adjust the direction of polarization of the incident light along the *y*-direction before it entered the beam-splitting prism. A dark field module (model) was used to project the incident light field onto a 100× Olympus objective that was corrected for high numerical aperture (NA = 0.9) and infinite conjugate and incident at a large angle to the sample surface. The diffracted light from the sample was collected by the same objective lens and directed into a 4f optical system, comprising three lenses and a pinhole. Finally, the Fourier plane of the diffracted light was projected via beam splitter BS2 onto CCD (QImaging Retiga R1 from Cairn Research) and fiber spectrometer (Exemplar®Plus, BTC655N). We used fiber with a core diameter of 100 $\mu m$ (model) to collect spectral signals displayed as spots on the Fourier plane. The collection end of the fiber was fixed on a support rod connected to an electric displacement platform, enabling point-by-point scanning with a 0.1 mm step using an externally-drive system to achieve dispersion imaging of the sample in momentum space. This configuration effectively eliminated scattered light interference from other positions around the sample, ensuring precise and focused light collection to produce sharp resonance peaks. Converting L3 into a lens with a focal length twice that of L3 allows for fast switching between momentum space and real space imaging. The ARPL measurements were performed with the same optical setup, using a 980 nm laser to vertically excite a single NC. After the PL signal of the sample is passed through an 800 nm short pass filter (placed in front of the L1), the Fourier image and the ARPL spectrum of the signal are imaged simultaneously using an FBP device.

**3.2 Micro-area measurement of upconversion lifetime for single-NCs**

The time-resolved confocal microscope (MicroTime 200, PicoQuant) was used to capture the fluorescence lifetime of individual upconversion nanocrystals (UCNCs). A picosecond pulsed diode laser (LDH-D-C-980, PicoQuant) with a wavelength of 980 nm was used as an excitation source and controlled via a programmable laser driver (PDL 828 "Sepia II", PicoQuant). The excitation radiation was transferred into an oil-immersion objective lens (100×, NA = 1.4, Olympus) via a dichroic mirror and concentrated onto the sample mounted in a high-precision piezoelectric scan stage. The upconversion luminescence (UCL) emitted from the sample was collected by the same objective and passed through a band-pass filter (center wavelength: 582 nm, Full Width at half maximum:75 nm) to effectively remove the residual excitation light. The filtered signal was then focused onto a confocal pinhole to ensure spatial resolution and directed to a Single Photon Avalanche Photodiode (SPAD, PMA Hybrid, PicoQuant). The detected photon events were time-tagged and analyzed by a time-correlated single-photon counting (TCSPC) module (HydraHarp 400, PicoQuant), enabling high-resolution fluorescence lifetime measurements. The MicroTime 200 system was operated under SymPho Time 64 software, which provides lifetime fitting and photon correlation analysis. This configuration allows for spatially resolved lifetime imaging and spectral characterization of individual nanocrystals, enabling a detailed analysis of their local photophysical properties under up-conversion excitation.

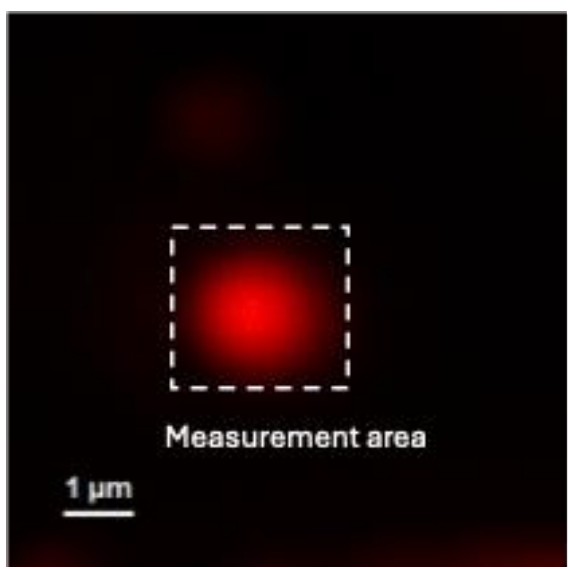

**Supplementary Fig. S9.** Confocal image of bright emission from a single-NC filled in a TPC under 0.2 MW power excitation at 976 nm.

# 4. TPC high-Q multi-BIC mode emission cone overlaps with upconversion emission cone on single-NC in cavity

As shown in the schematic diagram in Fig. S8b, the TPC is excited by a halogen lamp at a large incident angle using a 100× dark-field objective lens (numerical aperture NA = 0.9). The TPC's momentum-space emission image is captured by an infinite-corrected 4$f$ optical system equipped with a 532 nm bandpass filter. The momentum space equal-frequency contours are shown in Fig. S10a. The Fourier transform of the back-focus plane image reveals a horizontal slice of the high-Q multimode BIC mode emission cone at the 532 nm emission wavelength. Four mode arcs are observed, corresponding to the wave vectors of the square lattice, with their intersection points marking the M point of the topological plasmonic lattice.

Using the optical setup shown in Fig. S8a,c, single-NC in the TPC was pumped with 980 nm incident light. The coupled polariton emission cone was captured on the Fourier back focal plane image. The emission cone has similar momentum space distribution characteristics to the multimode BIC emission cone of the TPC.

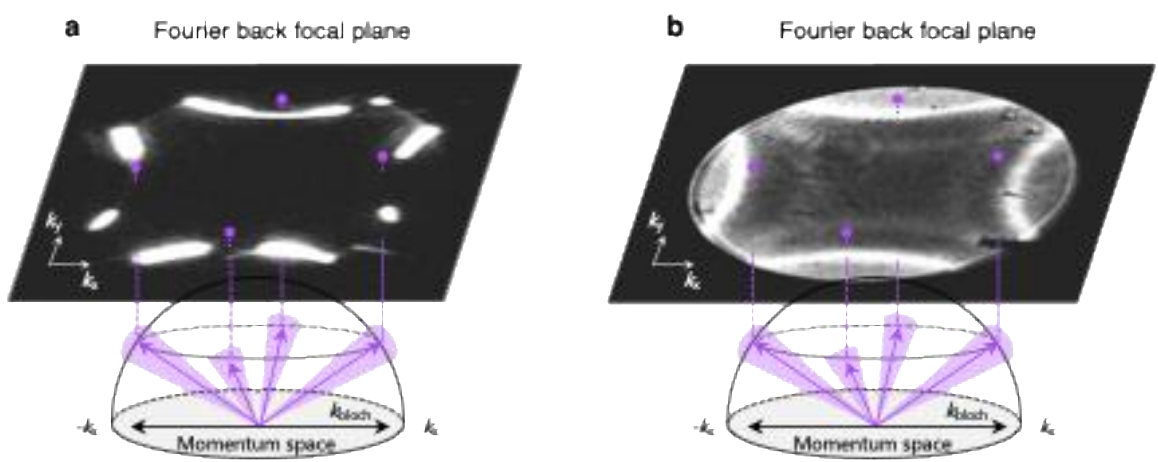

**Supplementary Fig. S10. a,** FBP image of the TPC under large-angle dark-field illumination (NA = 0.9) excited by a halogen lamp. The momentum space emission signal filtered at 532 nm shows a high-Q multimode BIC radiation pattern. The four crescent-shaped arcs correspond to the diffraction Bloch modes of the square lattice, with their intersection points marking the M point in momentum space. **b,** FBP image of a single-NC coupled to the TPC under

980 nm excitation. The observed emission cone exhibits a similar feature distribution to the intrinsic multimode BIC profile of the TPC, indicating overlap between the NC and the underlying multi-BIC modes.

## 5. Extraction of the Purcell factor of single-NCs with different particle sizes from the change in upconversion lifetime

The UCL dynamics in sensitizer-activator co-doped NCs under weak excitation are governed by the following rate equations:

$$\frac{d}{dt}n_{S_1} = \varphi_0 n_{S_0} - \Gamma_r^S n_{S_1} - (\omega_0 n_{A_0} + \omega_1 n_{A_0})n_{S_1} \tag{S1}$$

$$\frac{d}{dt}n_{A_1} = \omega_0 n_{S_1} n_{A_0} - \omega_1 n_{S_1} n_{A_1} - n_{A_1}/\tau_1^A \tag{S2}$$

$$\frac{d}{dt}n_{A_2} = \omega_1 n_{S_1} n_{A_1} - (\Gamma_r^A + \Gamma_{nr}^A)n_{A_2} \tag{S3}$$

where $n_{S_i}$ (in m$^{-3}$) and $n_{A_i}$ (in m$^{-3}$) denote population densities of sensitizer (e.g., Yb$^{3+}$) and activator (e.g., Er$^{3+}$) energy states $\varphi_0 = \sigma_S \rho/h\nu$ is the sensitizer absorption rate ($\sigma_S$:absorption cross-section, $\rho$: excitation power density) $\Gamma_r^S$, $\Gamma_r^A$ are radiative decay rates, $\Gamma_{nr}^A$ is the nonradiative decay rate $\omega_0, \omega_1$ represent Yb→Er energy transfer coefficients

The UCL decay is bottlenecked by the Yb$^{3+}$ energy level $^2F_{5/2}$ state relaxation:

$$\tau_0^{Yb} = (\Gamma_r^{Yb} + w_0 n_A)^{-1} \tag{S4}$$

The down-conversion luminescence (DCL) lifetime reflects the activator's metastable state decay:

$$\tau_0^A = (\Gamma_r^A + \Gamma_{nr}^A)^{-1} \tag{S5}$$

When coupled with a nanocavity, $\tau_0^{Yb}$ and $\tau_0^{Er}$ will be shortened by the nanocavity to be

$$\tau_{cav}^{Yb} = (F_P^{980} \Gamma_r^{Yb} + w_0 n_A)^{-1} \tag{S6}$$

$$\tau_{cav}^{Er} = (F_P^{\lambda_{em}} \Gamma_r^{Er} + \Gamma_{nr}^{Er})^{-1} \tag{S7}$$

The decay dynamics of UCL are primarily governed by the Yb$^{3+}$ sensitizer lifetime. That is, the measured UCL lifetimes approximate the Yb$^{3+}$ decay time:

$$\tau_0^U \approx \tau_0^{Yb}, \quad \tau_{cav}^U \approx \tau_{cav}^{Yb} \tag{S8}$$

according to Eq. (S8), we can express $F_P^{980}$ with $\tau_{cav}^U$ as

$$F_P^{980} = [(\tau_{cav}^U)^{-1} - w_0 n_{Er}]/\Gamma_r^{Yb}, \tag{S9}$$

Noting that $w_0 n_{Er}$ is extracted from the measured $\tau_0^U$ via $w_0 n_{Er} = (\tau_0^U)^{-1} - \Gamma_r^{Yb}$, we may also express $F_P^{980}$ as

$$F_P^{980} = \left[(\tau_{cav}^U)^{-1} - (\tau_0^U)^{-1}\right]/\Gamma_r^{Yb} + 1. \tag{S10}$$

Therefore, we can extract $F_P^{980}$ from the measured UCL lifetime change.

## 6. Topological properties of plasmonic lattice band

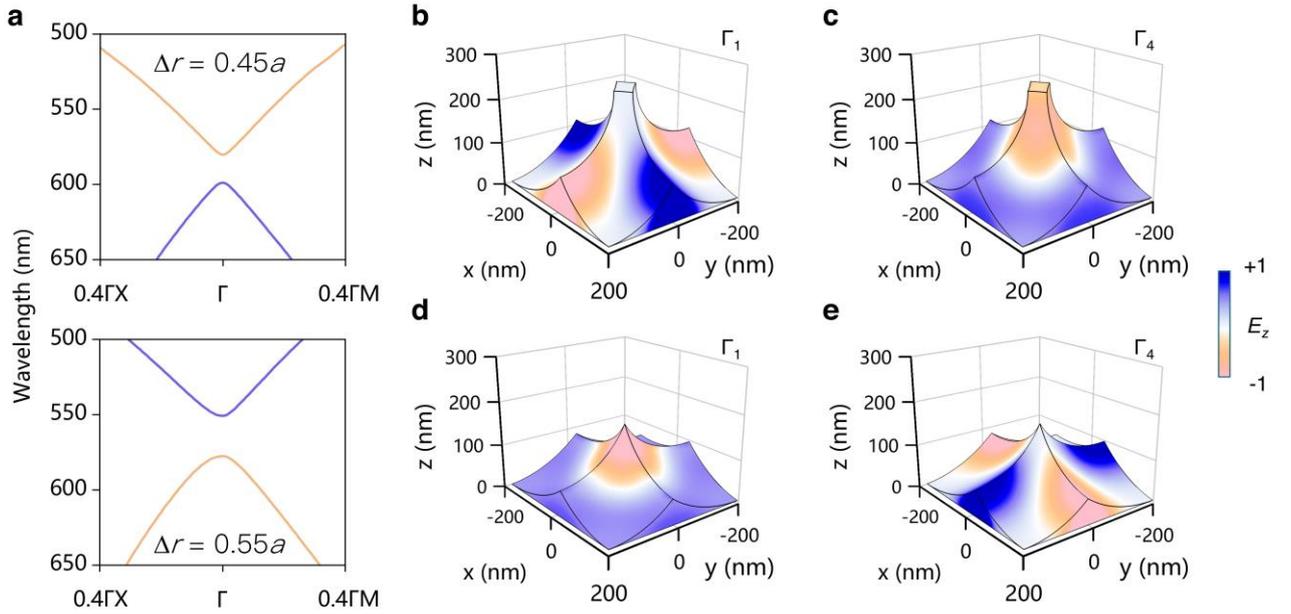

**Supplementary Fig. S11. Zak phase transition and modal inversion in plasmonic lattices. a,** Calculated 2D Zak phases of the lower and upper bands in the uncompressed (top) and compressed (bottom) plasmonic lattices. **b,c,** $E_z$ field distributions at the Γ point in the uncompressed lattice show a quadrupole mode (Γ$_1$) in the lower band and a monopole mode (Γ$_4$) in the upper band. **d,e,** Corresponding modal profiles in the compressed lattice exhibit inversion of these modes, confirming the topological phase transition.

We briefly discuss the band topology properties of plasmonic lattices after introducing sigma symmetry breaking. Considering the evolution of the unit cell from uncompressed ($\Delta r < 0.5a$) to compressed ($\Delta r > 0.5a$) as shown in Figure 1c of the main text, we observe that the accidental intersection points in the band structure undergo a process of closing and reopening (Figs. 2b,c), which is similar to the topological engineering of band structures in one-dimensional dielectric photonic crystals. To reveal the topological properties of the plasmonic lattice's energy bands, we theoretically investigated topological band inversion in the system. By gradually varying the lattice compression factor $\Delta r$, we observed topological band inversion, corresponding to a geometric phase shift in the mixed energy bands (Supplementary Fig. S11). This topological phase transition induced by $\sigma_h$ symmetry breaking can be characterized by the two-dimensional Zak phase, defined as follows:

$$\theta_\alpha = 2\pi \mathbf{P}_\alpha = \frac{1}{2\pi} \int_{FBZ} \mathrm{Tr}\left[A_\alpha(\mathbf{k})\right] d\mathbf{k}, \quad \alpha = x, y, \tag{S11}$$

where $\mathbf{P} = [Px, Py]$ indicates the 2D polarization of the lattice, $A_a(\mathbf{k}) = i \langle \psi_{n,k} | \partial_{k_a} | \psi_{n,k} \rangle$ is the Berry connection, and $|\psi_k\rangle$ denotes the Bloch eigenfunction on the $n$th band with wave vector $k$. We chose the $\Gamma$ point of the plasmonic lattice as the inversion center to calculate the two-dimensional Zak phases, as shown in Fig. S11a. In the uncompressed lattice (top panel), the lower and upper bands are characterized by Zak phases of ($\pi$, $\pi$) and (0, 0), respectively. Upon compression (bottom panel), a clear topological transition is observed: the Zak phases of the lower and upper bands switch to (0, 0) and ($\pi$, $\pi$), respectively, confirming a band inversion. The topologically nontrivial bands are highlighted by purple lines (Fig. S11a). At the critical state (Fig. 2b in the main text), the bandgap closes and the Zak phase becomes ill-defined, consistent with a topological phase transition. This transition is further corroborated by the observed modal inversion. As shown in Figs. S11b and S11c, the 3D electric field distributions ($E_z$) at the $\Gamma$ point in the uncompressed lattice reveal a quadrupolar mode in the lower band ($\Gamma_1$) and a monopolar mode in the upper band ($\Gamma_4$). In contrast, the compressed lattice exhibits an inversion of these modal profiles (Figs. S11d and S11e), further confirming the band inversion.

Notably, the q-BIC mode in the plasmonic lattice consistently appears in the band hosting the quadrupolar mode (Fig. S12). The observed band inversion indicates that the in-plane Bloch wave of the plasmonic lattice accumulates an additional geometric phase as the system parameters are tuned continuously. This behavior arises from strong coupling between the Bloch wave and LSPR, giving rise to plasmonic lattice resonances.

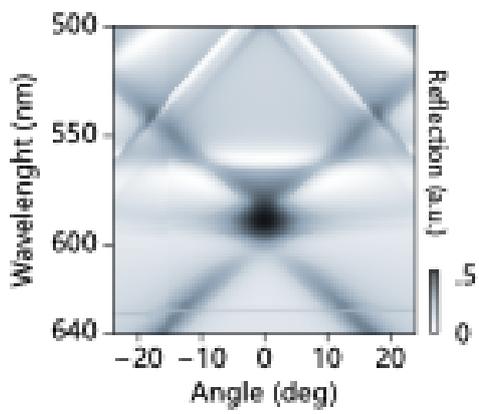

**Supplementary Fig. S12.** Angle-resolved reflection spectrum of an uncompressed lattice ($\Delta r = 0.45a$)

## 7. Single-NC upconversion far-field radiation direction comparison inside and outside the topological plasmonic cavity

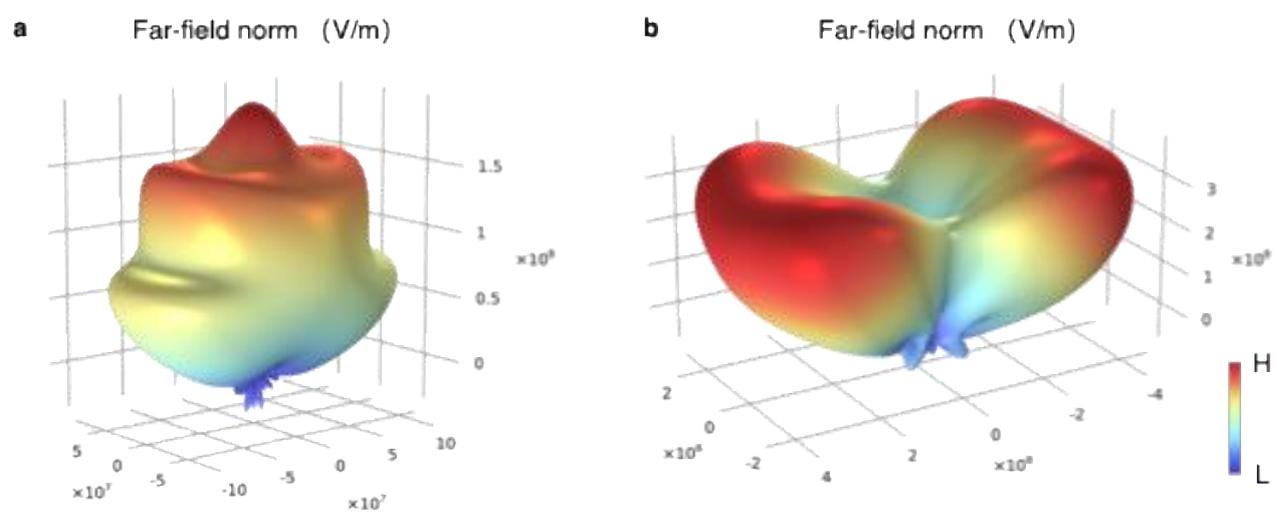

**Supplementary Fig. S13.** Comparison of far-field radiation patterns for single-NC upconversion emission simulated in COMSOL. A dipole source oriented along the *x*-axis was used to model the emission from a single-NC **a,** Radiation pattern of the NC placed directly on a flat aluminum film, representing the uncoupled case. The emission is broadly distributed in multiple directions with no clear confinement, indicating inefficient free-space coupling. **b,** Radiation pattern of the same NC

embedded within a topological plasmonic cavity (TPC). A highly directional emission profile is observed, concentrated into lobes orthogonal to the dipole orientation, consistent with coupling to finite Q multi-BIC modes supported by the TPC. The color scale represents the normalized electric field strength in the far field (*V/m*), with H and L denoting high and low intensity, respectively.

## 8. Emission Characteristics of a Single UCNP on a Flat Aluminum Film

To establish a baseline for evaluating the beam-shaping performance of the topological plasmonic cavity, we first characterized the emission properties of an isolated upconversion nanoparticle (UCNP) in the absence of any photonic lattice. A single UCNP (approximately 400 nm in diameter) was dispersed on a flat, unpatterned aluminum film. Fig. S14a shows the Fourier-plane (momentum-space) image of the upconversion luminescence (UCL) under continuous-wave excitation at 980 nm. The emission displays a nearly isotropic distribution, indicating that without the topological cavity, the emitted light lacks any spatial confinement or directional preference.

The corresponding ARPL spectrum along the high-symmetry direction (X–Γ–X) is presented in Fig. S14b. Three characteristic emission peaks of $Er^{3+}$ ions are clearly identified at 525 nm ($^4F_{7/2} \rightarrow {}^4I_{15/2}$), 550 nm ($^4S_{3/2} \rightarrow {}^4I_{15/2}$), and 650 nm ($^4F_{9/2} \rightarrow {}^4I_{15/2}$). These emissions are distributed across a broad angular range from −60° to +60°, with the maximum intensity observed around ±25°. This angular profile is consistent with electric dipole-type radiation on a metallic surface and confirms the presence of only low-Q emission channels in the absence of the topological BIC mode.

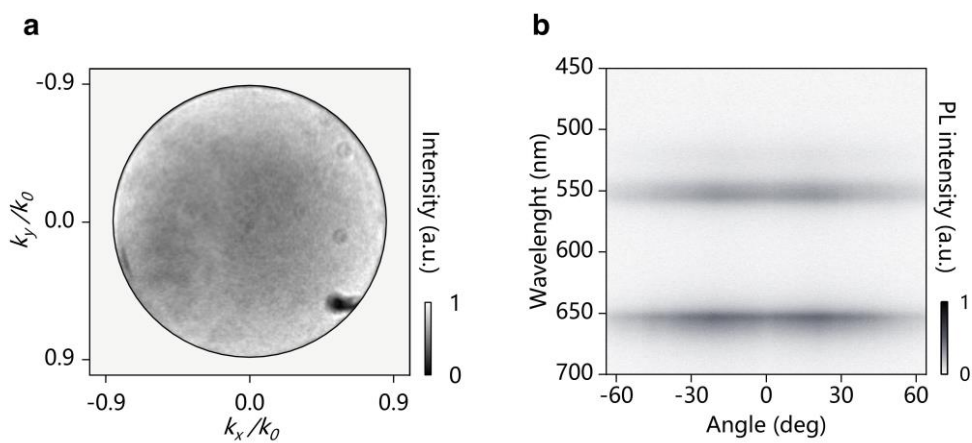

**Supplementary Fig. S14. a**, Fourier-plane image of the upconversion luminescence (UCL) from a single 400 *nm* upconversion NC placed on a flat aluminum film. The isotropic angular distribution confirms the absence of photonic

confinement or directional coupling. **b**, Angle-resolved PL spectrum of the same NC along the high-symmetry X–Γ–X direction, showing three emission peaks (525 *nm*, 550 *nm*, 650 *nm*) attributed to $Er^{3+}$ transitions. Emission extends broadly (±60°), consistent with dipolar radiation in free space.